\documentclass[%
 reprint,
amsmath,amssymb,
aps,
pra,
floatfix,
]{revtex4-2}

\usepackage{float}
\usepackage{natbib}

\usepackage{graphicx}
\usepackage{dcolumn}
\usepackage{bm}
\usepackage{xcolor}

\usepackage{braket}
\graphicspath{ {./images/} }

\begin{document}

\preprint{APS/123-QED}

\title{Squeezed light from a semiconductor amplifier}

\author{Andrey S. Aleksandrov}
\affiliation{Skolkovo Institute of Science and Technology, Russia, 121205 Moscow, Skolkovo, Bolshoi Bulvar 30, 1}
\affiliation{Russian Quantum Center, Russia, 121205 Moscow, Skolkovo, Bolshoi Bulvar 30, 1}

\author{Anatoly V. Masalov}%
\affiliation{Russian Quantum Center, Russia, 121205 Moscow, Skolkovo, Bolshoi Bulvar 30, 1}
\affiliation{P.N. Lebedev Physical Institute, Russia, 119991 Moscow, Leninskiy Prospect 53}

\date{\today}

\begin{abstract}
    The possibility of generating light in a squeezed quantum state during saturated light amplification in a semiconductor amplifier is theoretically analyzed. Due to four-wave mixing process during amplification, weak components of the spontaneous emission of the medium with symmetric frequency shifts relative to the strong amplified wave acquire the quantum properties of two-mode squeezed light. The squeezing of quadrature components occurs in the range of frequency shifts of several GHz ($\sim$ the inverse relaxation time of the charge carrier density) and can reach 10 dB or more under optimal conditions. The formation of a two-mode squeezing phase facilitates the observation of the squeezing effect, in which a strong wave, playing the role of a local wave, exhibits quadrature noise suppression without additional homodyne detection.
\end{abstract}

\maketitle

\section{Introduction}

The diversity of semiconductor media is attractive for implementing nonlinear optical transformations of radiation. These media are characterized by nonlinear susceptibility $\chi^{(3)}$ both in the transparency region and in the absorption band. In the transparency region, the nonlinearity of semiconductor media is similar to that of traditional dielectric media - crystals and gases: the nonlinearity is virtually inertialess and manifests itself at relatively high radiation intensities ($>10^9 \text{ W/cm}^2$). In the absorption band, the nonlinear properties of semiconductor media are also similar to those of atomic-molecular media with resonant nonlinearity, where absorption (or gain) under the action of light depends on the radiation intensity. Under such conditions, the nonlinearity is significantly inertial ($\sim 10^{-9}$ s) and manifests itself at significantly lower radiation intensities ($I_{\text{sat}} \sim 10^7 \text{ W/cm}^2$).

The cubic nonlinearity of media ($\chi^{(3)}$-susceptibility) is attractive for generating light in a squeezed quantum state using the four-wave mixing (FWM) scheme. The authors of \cite{fox1995quadrature, schucan1998femtosecond} used the $\chi^{(3)}$ nonlinearity in ZnSe and CdSe in the transparency region to generate squeezed light; relatively low degrees of squeezing of 0.3-0.4 dB (0.7 dB is inferred at the crystal) were achieved, and high intensities of fs laser pulses were required. As theoretical calculations show, the potential of semiconductor media for generating squeezed light is not limited to 1 dB. For the $\text{Al}_x\text{Ga}_{1-x}\text{As}$ waveguide medium, the theoretical prediction yielded an 8.2 dB quadrature-noise reduction, limited only by the self-phase modulation of the pump \cite{ho1995single}.

\begin{figure}
    \centering
    \includegraphics[width=0.43\textwidth]{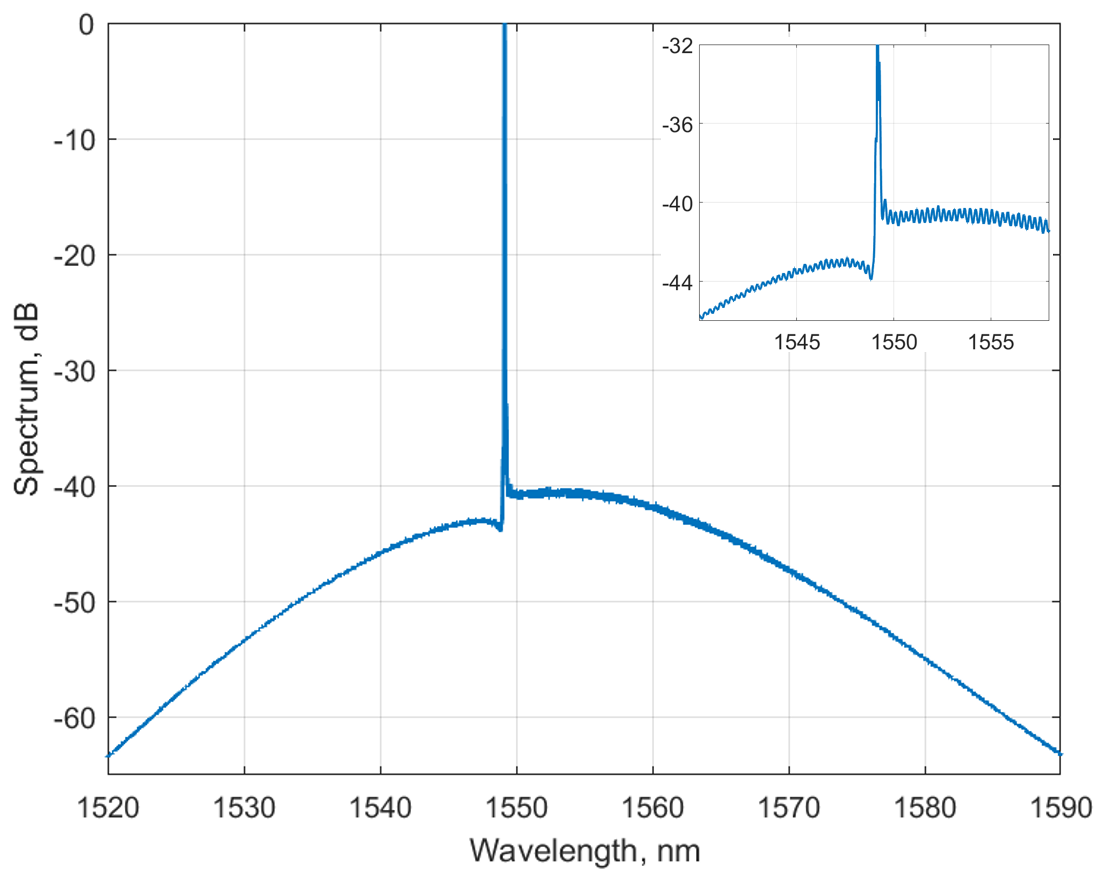}
    \caption{The mode of single-frequency radiation against the background of multiple subthreshold modes (nearly unresolved) of an InP laser. Insert: enlarged fragment of the spectrum. Courtesy of A. N. Danilin.}
    \label{fig:Dan}
\end{figure}

In the absorption region, the nonlinearity of semiconductor media has been well studied thanks to theoretical and experimental studies of the operating regimes of semiconductor lasers \cite{agrawal2013semiconductor}. Here, the nonlinearity manifests itself as a cubic nonlinearity only at low radiation intensities ($\ll I_{\text{sat}}$). As the intensity increases, when the charge carrier density changes significantly, the  balance equations are applicable in the description. In laser semiconductor media, the spectral absorption/gain band significantly exceeds the spectral widths of the acting radiation. This allows to use balance equations in the limit of zero transverse relaxation time. Within this model, qubic nonlinearity and four-wave mixing process in semiconductor media are developed with intensity dependent charge carrier density. A bright manifestation of four-wave mixing in a gain laser medium is the "Bogatov effect": a decrease and increase in the gain of a weak waves in the presence of a strong one, depending on the sign of the frequency detuning of the weak wave \cite{bogatov1975anomalous, agrawal1987fourwave, ogasawara1988longitud, ishikawa1981longitud}. An illustration of this effect is the shape of the spectrum of subthreshold modes of a single-frequency semiconductor laser \cite{galiev2018spectrum}. Weak red-shifted subthreshold modes gain some advantage over blue-shifted ones (Fig.\ref{fig:Dan}). The effect takes place at frequency shifts within band limited by the inverse of charge carrier relaxation time ($\sim 10^{-9}$ s). The observed four-wave mixing mechanism in a gain medium can be used to generate light in a squeezed quantum state, just as has been demonstrated with traditional dielectric media \cite{fox1995quadrature, schucan1998femtosecond}. The proposed FWM process in a semiconductor amplifier is analogous to FWM in Kerr-based microresonators where squeezing levels of 1.0–1.7 dB have been demonstrated \cite{Dutt2015, Vaidya2020, Zhao2020} and where the low efficiency of Kerr nonlinearity is overcome through the buildup of radiation within the resonator.

This paper presents theoretical analysis of the characteristics of squeezed light generated during strong wave propagation through an amplifying semiconductor medium. Two-mode squeezed light components appear as symmetric side-bands around strong wave. Initially, weak side-band components of the radiation are inevitably present at the first layers of amplifying medium as spontaneous emission. Suppression of quantum fluctuations of weak side-band components due to FWM with a strong wave was subjected to calculation. The appeal of this squeezed light generation scheme is that the strong wave can serve as a local wave when recording the degree of squeezing.

\section{Saturated gain}
The quantitative description of squeezed light formation during amplification is based on the standard equations relating light and carrier density of amplifying medium:
\begin{align}
    \frac{dA}{dz} & = \frac{1 - i\alpha}{2}\sigma nA, \label{eq:Adz}
    \\
    \frac{dn}{dt} & = p - \frac{n}{\tau_e} - 2\sigma n\frac{I}{\hbar\omega}, \label{eq:Ndt}
\end{align}
where:\\
$A$ and $I$ are the complex amplitude and intensity of the light wave, $|A|^2=I$;\\
$n$ is the inversion of carrier density, providing gain of radiation;\\
$p$ is the pumping power (proportional to the pump current);\\
$\tau_e$ is the relaxation time of carrier density (in semiconductor media, $\tau_e \sim 10^{-9}$ s);\\
$\alpha$ is the Henry factor, which characterizes the change in the refractive index of a semiconductor medium with a change in inversion of carrier density.\\
The equation (\ref{eq:Adz}) describes the amplification of the complex amplitude of a light wave $A$ as it propagates in a medium along the $z$ coordinate. The equation (\ref{eq:Ndt}) defines a local relationship between inversion $n$ and light intensity $I$.

The equations (\ref{eq:Adz}) and (\ref{eq:Ndt}) have the form of standard balanced equations of a two-level amplifying medium, where the relationship between gain and inversion is given through the stimulated emission/absorption cross section $\sigma$. In semiconductor media, the relationship between gain and inversion is described by other quantities, but the proportionality of gain and inversion is preserved. Introduction of the saturation intensity $I_{\text{sat}}=\hbar\omega/(2\sigma\tau_e)$ makes the equation (\ref{eq:Ndt}) universal:
\begin{equation}
    \frac{dn}{dt} = p - \frac{n}{\tau_e}\left(1 + \frac{I}{I_{\text{sat}}}\right).
    \label{eq:Ndt1}
\end{equation}
For numerical calculations, it is convenient to normalize the inversion $n$ by the value of $p\tau_e$. Then, denoting $N=n/(p\tau_e)$ and $S=I/I_{\text{sat}}$, we get the normalized form of the equation (\ref{eq:Ndt1}):
\begin{equation}
    \tau_e\frac{dN}{dt} = 1 - N\left(1 + S\right). \label{eq:NdtN}
\end{equation}
The use of this normalization in the equation (\ref{eq:Adz}) leads to its modification:
\begin{equation}
    \frac{dA}{dZ} = \frac{1 - i\alpha}{2}NA, \label{eq:AdzN}
\end{equation}
where $Z=zp\tau_e\sigma$ and we may still consider $|A|^2=S$.

When the radiation intensity is low and does not saturate the inversion, the intensity increases as $S(Z) = S_0e^Z$, where the normalized coordinate $Z$ defines the increment of the unsaturated gain of the medium.

The initial radiation field at the entrance to the medium will be represented as the sum of a strong wave $Ae^{-i\omega t}$ with two weak side-bands with symmetrically shifted frequencies $a_+e^{-i(\omega+\Omega)t}$ and $a_-e^{-i(\omega -\Omega)t}$:
\begin{align}
    E = e^{-i\omega t}\left( A + a_+e^{-i\Omega t} + a_-e^{i\Omega t}\right),
    \label{eq:Eini}
\end{align}
with following intensity:
\begin{align}
    |E|^2= S + \Delta S \approx S & + (a_+A^* + a_-^*A)e^{-i\Omega t} \nonumber\\
    & + (a_+^*A+a_-A^*)e^{i\Omega t}.
    \label{eq:Sini}
\end{align}

The side-band modes originate as  spontaneous radiation of the input layers of amplifying medium.

During the amplification process, the inversion of carrier density $N$ acquires oscillations at the beat frequency $\Omega $:
\begin{equation}
    N \rightarrow N + \Delta N \approx N + \delta Ne^{-i\Omega t} + \delta N^*e^{i\Omega t}.
    \label{eq:Nini}
\end{equation}
Equations (\ref{eq:NdtN}) and (\ref{eq:AdzN}) can be solved under the assumption of weak fields of the modulation components. In the zero approximation, we have a stationary equation for the inversion $N$ and the field of the main wave $A$:
\begin{align}
    \frac{dA}{dZ} & = \frac{1-i\alpha}{2}\frac{A}{1 + |A|^2},    \label{eq:Adz0}
    \\
    \frac{dS}{dZ} & = \frac{S}{1 + S},
    \label{eq:Sdz0}
    \\
    N & = \frac{1}{1+S},
    \label{eq:Ndz0}
\end{align}

The solution of the equations (\ref{eq:Adz0}-\ref{eq:Ndz0}) has the form:
\begin{align}
    Z & = S(Z) - S_0 + \ln{\frac{S(Z)}{S_0}},
    \\
    A(Z) & = A_0\exp \left(\frac{1-i\alpha}{2}\ln{\frac{S(Z)}{S_0}} \right),
\label{eq:SandA}
\end{align}

In the first approximation, substituting (eq. \ref{eq:Eini}) and (eq. \ref{eq:Nini}) in (eq. \ref{eq:NdtN}) and (eq. \ref{eq:AdzN}), we get
\begin{align}
    \tau_e\frac{d}{dt}\Delta N \approx & - (1 + S)\Delta N - N\Delta S,
    \label{eq:NdtDelta}\\
    \frac{d}{dZ}\Delta A \approx & \frac{1-i\alpha}{2}(N\Delta A + A\Delta N).
    \label{eq:AdzDelta}
\end{align}
Collecting the oscillations at the frequencies $\pm\Omega$, we have the expression for inversion oscillations $\delta N$:
\begin{equation}
    \delta N = - \frac{a_+A^*+a_-^*A}{1+S}\frac{1}{1 + S - i\Omega \tau_e},
    \label{eq:DNsol}
\end{equation}
and the equations for weak fields
\begin{align}
    \frac{da_+}{dZ} = & \frac{1-i\alpha}{2(1+S)}\left\{a_+ -A \frac{a_+A^*+a_-^*A}{1+S - i\Omega \tau_e}\right\},
    \label{eq:Da+}
    \\
    \frac{da_-}{dZ} = & \frac{1-i\alpha}{2(1+S)}\left\{a_- - A\frac{a_+^*A+a_-A^*}{1+S+i\Omega \tau_e}\right\}.
    \label{eq:Da-}
\end{align}
Introduction of relative amplitudes $b_+(Z)=a_+(Z)/A(Z)$ and $b_-(Z)=a_-(Z)/A(Z)$ simplifies the equations:
\begin{align}
    \frac{db_+}{dZ} = & -\frac{1-i\alpha}{2}\frac{b_+ + b_-^*}{1+S-i\Omega \tau_e}\frac{S}{1+S},
    \label{eq:Db+}
    \\
    \frac{db_-}{dZ} = & -\frac{1-i\alpha}{2} \frac{b_+^* + b_-}{1+S+i\Omega \tau_e} \frac{S}{1+S},
    \label{eq:Db-}
\end{align}
and, taking into account $dZ=dS\frac{1+S}{S}$ (see eq. \ref{eq:Sdz0}), gives solutions
\begin{align}
    b_+(Z) = b_+(0) - \frac{1-i\alpha}{2}(b_+(0) + b_-^*(0))\frac{S(Z)-S_0}{1+S(Z)-i\Omega \tau_e}
    \label{eq:b+sol}
    \\
    b_-(Z) = b_-(0) - \frac{1-i\alpha}{2}(b_+^*(0) + b_-(0))\frac{S(Z)-S_0}{1+S(Z)+i\Omega \tau_e}
    \label{eq:b-sol}
\end{align}
and also for the initial amplitudes
\begin{align}
    a_+(Z) = & K \left(a_+(0) -\frac{1-i\alpha}{2}\left(a_+(0) + a_-^*(0)\right)M(\Omega )\right),
    \label{eq:a+sol}
    \\
    a_-(Z) = & K \left(a_-(0) - \frac{1-i\alpha}{2}\left( a_-(0) + a_+^*(0) \right) M(-\Omega ) \right).
    \label{eq:a-sol}
\end{align}
Here $K=\exp\left(\frac{1-i\alpha}{2}\ln{\frac{S(Z)}{S_0}}\right)$ is the amplitude gain of the strong mode (see eq. \ref{eq:SandA}), where $|K|^2=S(Z)/S_0$ is the energy gain (saturated). The function
\begin{equation}
    M(\Omega)=\frac{S(Z)-S_0}{1+S(Z)-i\Omega \tau_e},
    \label{eq:M}
\end{equation}
is responsible for the FWM effect: $M(-\Omega)=M^*(\Omega)$. 
The role of FWM decreases both in the limit of low intensity of the strong wave, when $S\ll1$, and at high frequencies of $\Omega\gg (1+S)/\tau_e$, where  the carrier population does not follow oscillations of light intensity and FWM disappears. This limits the frequency range where FWM is relevant and where suppression of quantum fluctuations can be expected.

\begin{figure}
    \centering
\includegraphics[width=0.43\textwidth]{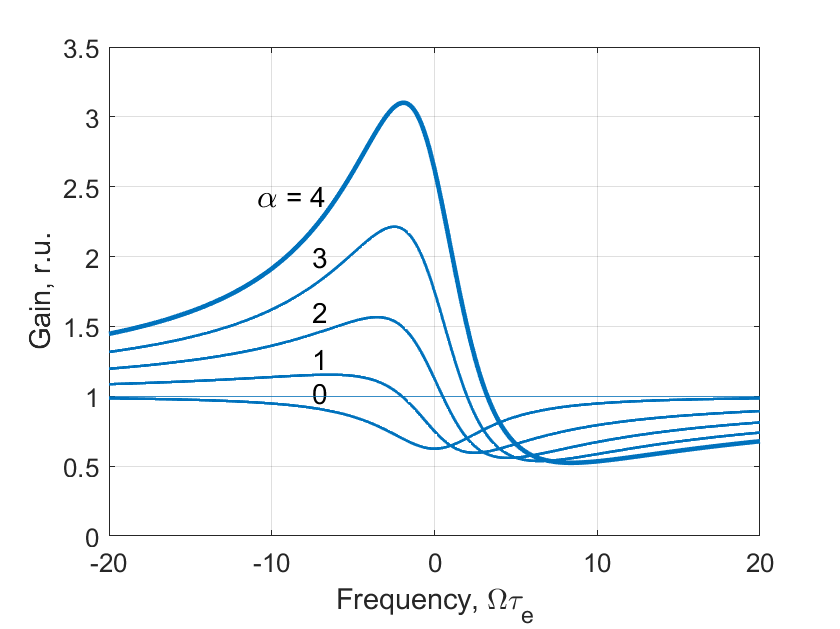}
    \caption{The spectrum of the relative energy gain of weak modes due to FWM for input intensity $S_0=1$, gain $Z=3$ and different values of Henry factor $\alpha$.}
    \label{fig:PM}
\end{figure}

The relationship of the weak modes $a_+$ with $a_-^*$ and $a_+^*$ with $a_-$ in (\ref{eq:a+sol}) and (\ref{eq:a-sol}) definitely indicates the possibility of squeezing of field fluctuations during homodyne detection of two side-band components with a local wave at the frequency of a strong field.
The equations (\ref{eq:a+sol}) and (\ref{eq:a-sol}) show that the gain coefficients of the modes on different sides of the frequency $\omega$ have different values. If we consider the initial amplitudes of the weak modes to be independent random seeds of the same power, then the energy of the weak modes at the output of the amplifier is characterized by the energy gain
 \begin{equation}
    |K|^2 \left( \left|1-\frac{1-i\alpha}{2}M(\Omega)\right|^2 + \left|\frac{1-i\alpha}{2}M(\Omega)\right|^2\right).
\label{eq:K}
\end{equation}
Positive $\Omega$ in (\ref{eq:K}) describes blue-shifted weak-mode amplification, while negative $\Omega$ corresponds to red-shifted mode. The spectrum of the energy gain of weak modes (eq. \ref{eq:K}) (without the common factor $|K|^2=S(Z)/S_0$) is shown in Fig.\ref{fig:PM} for different values of $\alpha$. Except for the case $\alpha=0$, the gain of the blue-shifted weak mode is lower than the total coefficient $|K|^2$, and the gain of the red-shifted mode is larger. 
It was precisely this effect that was observed by Bogatov et al \cite{bogatov1975anomalous} and illustrated in \cite{galiev2018spectrum}.  Fig.\ref{fig:PM} shows that the magnitude of the FWM effect noticeably increases with an increase in the Henry factor $\alpha$, which in semiconductor laser materials can reach a value of 4 - 6.  At low Henry factor $\alpha \ll 1$ the relationship of the weak modes in (\ref{eq:a+sol}) drops down and FWM is less pronounced. That is why in all further calculations of quadrature noise in the amplification process were performed at $\alpha =4$.

The relationship between the mode amplitudes and their complex conjugates in (eq. \ref{eq:a+sol}) and (eq. \ref{eq:a-sol}) clearly indicates the possibility of quadrature squeezing of the total field. We consider equations (\ref{eq:a+sol}) and (\ref{eq:a-sol}) as copies of quantum solutions for the operators $a_+\rightarrow\hat{a}_+$ and $a_-\rightarrow\hat{a}_-$ in the Heisenberg picture. Direct replacement of equations (\ref{eq:a+sol}) and (\ref{eq:a-sol}) with their quantum counterparts is valid when the commutators $[\hat{a}_+,\hat{a}_+^+]$ and $[\hat{a}_-,\hat{a}_-^+]$ are preserved. In equations (\ref{eq:a+sol}) and (\ref{eq:a-sol}) of the form $\hat{a}=c_1\hat{a}_{10}+c_2\hat{a}_{20}^+$, the commutator $[\hat{a},\hat{a}^+]$ is preserved under the condition $|c_1|^2-|c_2|^2=1$. 
In our case, this condition turns out to be fulfilled at $S\gg1$. In this case, for the two-mode field at the output of the amplifier (this is the actual field of two weak waves, not to be confused with (eq. \ref{eq:Eini}))
\begin{align}
E(Z)=\frac{1}{2}\bigl(&a_+(Z)e^{-i(\omega+\Omega)t}+a_+^*(Z)e^{i(\omega+\Omega)t} \nonumber \\
+&a_-(Z)e^{-i(\omega-\Omega)t}+a_-^*(Z)e^{i(\omega-\Omega)t}\bigr),
\end{align}
we may calculate $\braket{\Delta E^2}$ without changing the notation to quantum ones with initial conditions
\begin{equation}
    \braket{a_+(0)a_+^*(0)}=\braket{a_-(0)a_-^*(0)}=\frac{1}{2},
    \label{eq:fluct}
\end{equation}
\begin{widetext}
where all other second-order correlators are zero:
    \begin{align}
        \braket{\Delta E^2} & = \frac{1}{2}
        \left(\braket{a_+(Z)a_+^*(Z)} +
        \braket{a_-(Z)a_-^*(Z)} + \braket{a_+(Z)a_-(Z)} e^{-2i\omega t} + 
        \braket{a_+^*(Z)a_-^*(Z)} e^{2i\omega t}\right) =\nonumber\\
        & = \frac{|K|^2}{2} \left\{\braket{a_+(0) a_+^*(0)} \left| 1-\frac{1-i\alpha}{2}M\right|^2 + \braket{a_-(0) a_-^*(0)} \left| \frac{1-i\alpha}{2}M\right|^2
        \right\} + \nonumber\\
        & + \frac{|K|^2}{2} \left\{\braket{a_+(0) a_+^*(0)} \left| \frac{1-i\alpha}{2}M\right|^2 + \braket{a_-(0) a_-^*(0)} \left| 1 - \frac{1-i\alpha}{2}M^*\right|^2
        \right\} - \nonumber\\
        & - \left\{\frac{K^2}{2}\braket{a_+(0) a_+^*(0)} 
        \left( 1-\frac{1-i\alpha}{2}M\right) \frac{1-i\alpha}{2} M^*e^{-2i\omega t} + \text{c.c.}\right\} -\nonumber\\
        & - \left\{\frac{K^2}{2}\braket{a_-(0) a_-^*(0)} \left( 1-\frac{1-i\alpha}{2}M^*\right)\frac{1-i\alpha}{2}M e^{-2i\omega t} + \text{c.c.}\right\}.
        \end{align}
        After substitution (eq. \ref{eq:fluct}) we get
        \begin{align}
        \braket{\Delta E^2} & = \frac{|K|^2}{4} \left\{\left| 1-\frac{1-i\alpha}{2}M\right|^2 +\left| 1-\frac{1-i\alpha}{2}M^*\right|^2+ 2\left| \frac{1-i\alpha}{2}M\right|^2
        \right\} - \nonumber\\
        & - \left\{\frac{K^2}{4}\frac{1-i\alpha}{2} \left( M + M^* - \left( 1-i\alpha \right)|M|^2 \right)e^{-2i\omega t} + \text{c.c.}\right\}.
        \label{eq:Enoise}
    \end{align}
\end{widetext}
According to (eq. \ref{eq:Enoise}) field fluctuations contain a time-constant component $C$ and a component oscillating at twice the frequency $2\omega$ with an amplitude of $V$,
\begin{align}
    \braket{\Delta E^2} & =
    \frac{|K|^2}{4} C - 
    \left( \frac{K^2}{4} V e^{-2i\omega t} + \text{c.c.} \right) = \nonumber\\
    & = \frac{|K|^2}{4} \big( C - 2 |V|\cos\left( 2\omega t - \theta \right) \big),
    \label{eq:DE2}
\end{align}
where:
\begin{equation}
    C = (1+\alpha^2)|M|^2 - 2\text{Re}M +2,
    \label{eq:C}
\end{equation}
\begin{equation}
    V=(1-i\alpha)\left( \text{Re}M - \frac{1-i\alpha}{2}|M|^2\right),\ \theta = \arg (K^2V).
    \label{eq:V}
\end{equation}
The temporal variations of fluctuations in the two-mode field (eq. \ref{eq:DE2}) is the characteristic of squeezing. Obviously, the minimum fluctuations of the field (eq. \ref{eq:DE2}) have the value
\begin{equation}
    \braket{\Delta E^2}_{\min} = \frac{|K|^2}{4} \left( C - 2 |V|\right).
    \label{eq:DE2min}
\end{equation}
Calculations of $S(Z)$ and $M$ at given $Z$, $S_0$ (by (11)) and at given $\Omega \tau_e$ (by (23)) were used for estimation of squeezing degree and optimal parameter ranges. The largest FWM effect on squeezing during amplification should be expected in the range $S=I/I_{\text{sat}}\geq1$, where strong wave affects the inversion and the amplification becomes nonlinear. At lower input intensity  $S_0=I_0/I_{\text{sat}}\ll1$, $M\rightarrow 0$, the noise level of two-mode field (eq. \ref{eq:fluct}) obviously equals to the spontaneous noise, amplified by a phase-insensitive amplifier $|K|^2(\frac{1}{4}+\frac{1}{4})=|K|^2/2$, and squeezing is zero. Therefore, further calculations are presented for the ranges $S=I/I_{\text{sat}}\geq1$ and $Z\geq1$, where FWM is significant. In the next Section we present calculations of the minimal field fluctuations (eq. \ref{eq:DE2min}) in the form of the degree of squeezing 
\begin{equation}
    \text{Squeezing (dB) }= - 10\log\frac{\braket{\Delta E^2}_{\min}}{1/2},
    \label{eq:Squ}
\end{equation}
where field fluctuations are compared to $1/2$, i.e., 1/4 for each mode.

\section{Results} 
\begin{figure}
    \centering \includegraphics[width=0.43\textwidth]{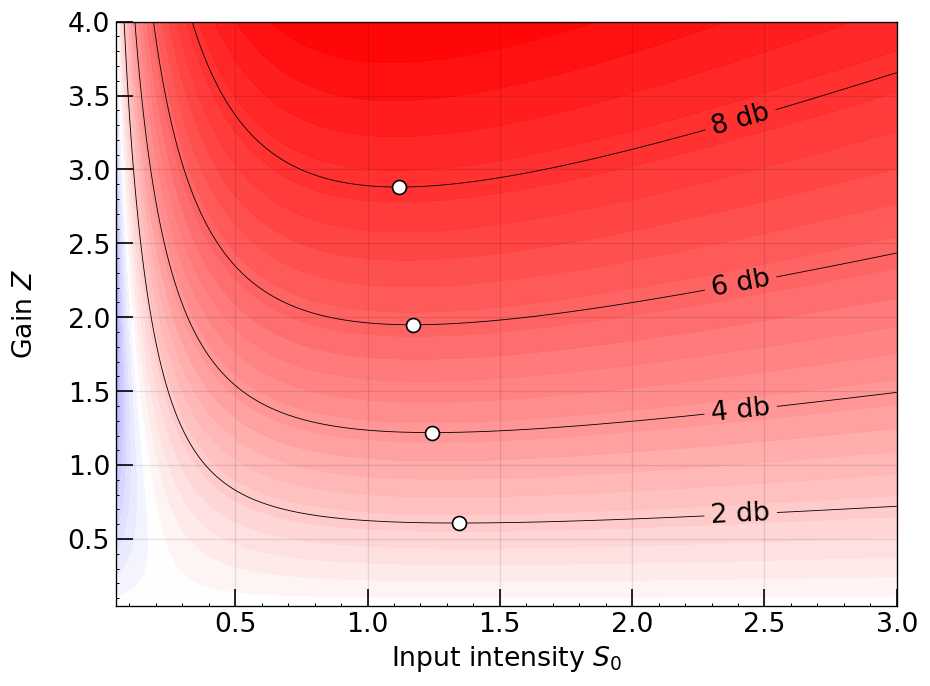}
    \caption{Degree of squeezing $\braket{\Delta E^2}_{\text{min}}$ depending on the input intensity $S_0$ and gain $Z$, frequency $\Omega\tau_e = 0.01$. The circles denote the values of the minimum gain $Z$ to achieve the specified squeezing.}
    \label{fig:SqvsZandS0}
\end{figure}

\begin{figure}
    \centering \includegraphics[width=0.43\textwidth]{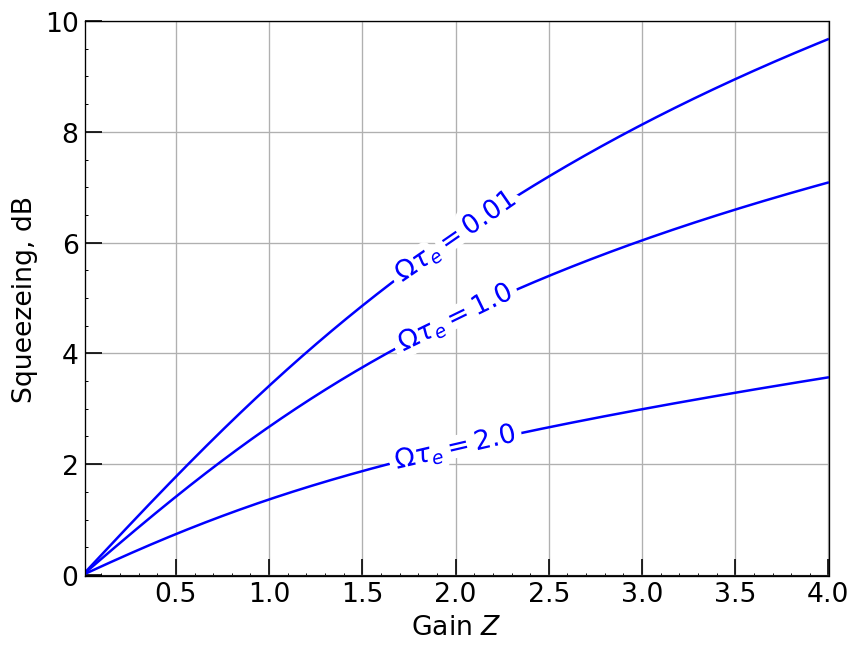}
    \caption{Degree of squeezing $\braket{\Delta E^2}_{\text{min}}$ depending on $Z$ at input intensity $S_0=1.5$ and various modulation frequencies $\Omega\tau_e$.}
\label{fig:SqvsZ}
\end{figure}

First of all, the degree of squeezing in our model of semiconductor amplifier was calculated depending on input light intensity $S_0$ and gain $Z$ (increment of unsaturated gain), Fig.\ref{fig:SqvsZandS0}. In further calculations we used the value of $\alpha =4$, which is typical for GaAs laser media \cite{henry1982theory}. For lower values of $\alpha$ squeezing is decreasing. In addition, a number of examples use low frequencies of $\Omega\tau_e\ll1$, which unambiguously increase the squeezing effect. Data on  Fig.\ref{fig:SqvsZandS0} show that squeezing reaches its maximum value in the range of input intensities $S_0 = 1-1.5$. This range can be considered optimal for squeezing. At the same time, the degree of squeezing increases monotonously with increasing gain $Z$, Fig.\ref{fig:SqvsZ}. It can be seen that the squeezing value can reach values attractive for experimental implementation: $\approx 10$ dB at $Z=4$ ($e^Z=55$) and at low frequencies of $\Omega\tau_e \sim 1$.

\begin{figure}
    \centering \includegraphics[width=0.43\textwidth]{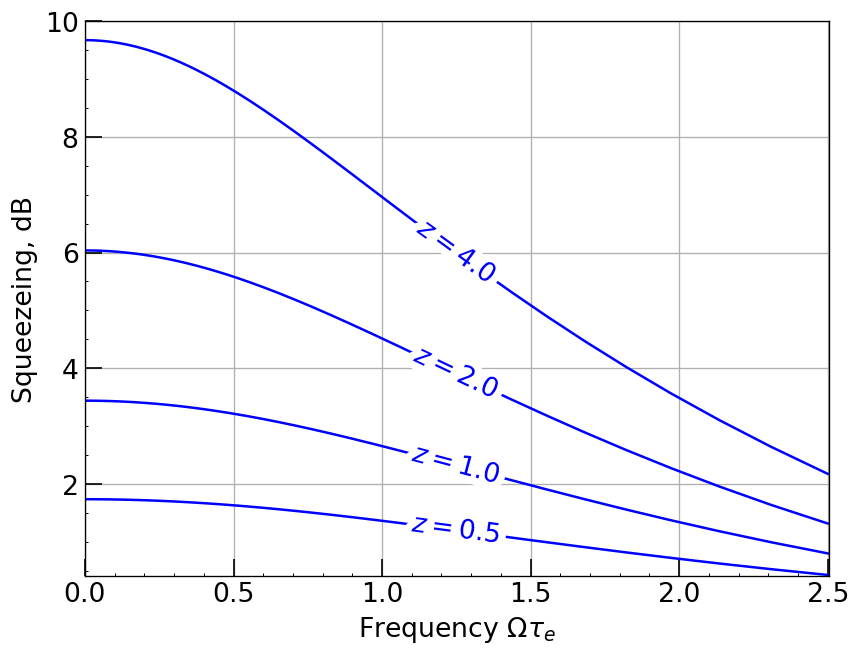}
    \caption{Degree of squeezing $\braket{\Delta E^2}_{\text{min}}$ depending on $\Omega\tau_e$ at different gain $Z$ for input intensity $S_0=1.5$.}
    \label{fig:SqvsO}
\end{figure}

\begin{figure}
    \centering \includegraphics[width=0.43\textwidth]{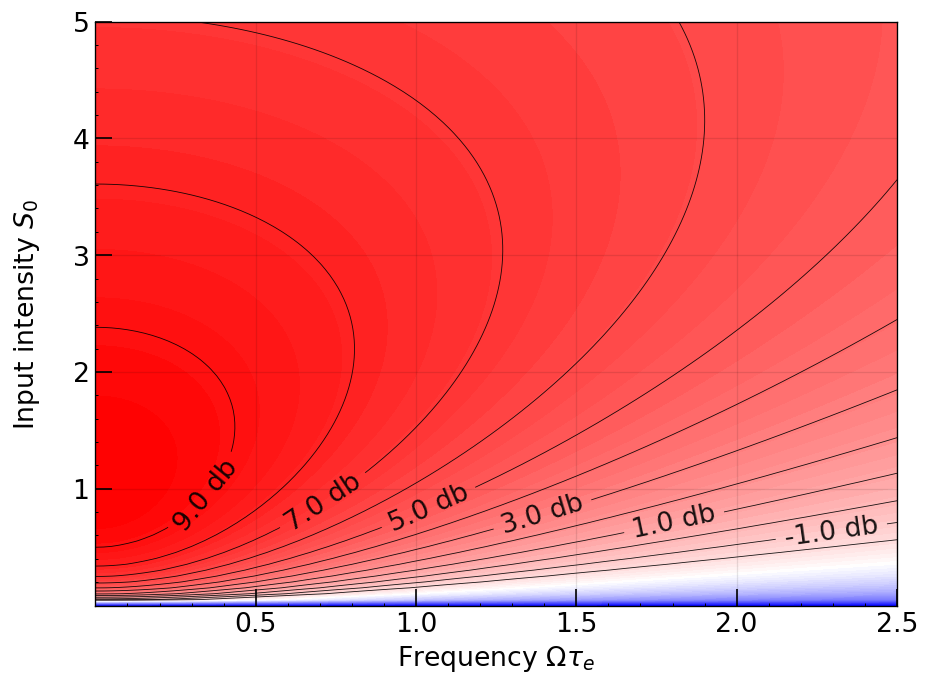}
    \caption{Degree of squeezing $\braket{\Delta E^2}_{\text{min}}$ depending on the frequency $\Omega\tau_e$ and input intensity $S_0$ for gain $Z=4$.}
    \label{fig:SqvsOandS0}
\end{figure}

At optimal input intensity the degree of squeezing is decreasing with growing frequency shift $\Omega\tau_e$, Fig.\ref{fig:SqvsO}. The full picture of squeezing depending on frequency shift $\Omega\tau_e$ at different input intensities is shown in Fig.\ref{fig:SqvsOandS0}. The range of significant squeezing extends up to $\Omega\tau_e\leq 1$.

The spectrum of light at the amplifier output consists of a strong wave above the background of multiple pairs of components of squeezed light with a continuous range of frequency offsets $\Omega$. In the process of direct photodetection the strong wave acts as a local wave for the components of two-mode squeezed light with fixed phase shift. With a favorable phase shift, direct detection reveals the degree of quadrature squeezing of weak field components. However, in our case of saturated amplification the relative phase shift is far from optimal for direct detection of squeezing. Calculations of the noise of oscillating amplitude $J_\Omega = A^*a_+(Z)+Aa_-^*(Z)$ in the direct detection of output intensity (eq. \ref{eq:Sini}) yield:
\begin{equation}
    \braket{\Delta J^2}=\frac{1}{2}|A|^4\left(C-2\text{Re} V\right).
    \label{eq:Exp}
\end{equation}
Estimates based on this formula indicate that direct photodetection reveals only a small fraction of squeezing or even anti-squeezing. Thus, measuring the degree of squeezing requires additional effort.

\section{Conclusions}
The presented estimations show the possibility of generating squeezed light when amplifying a strong quasi-monochromatic wave in a semiconductor amplifier by converting spontaneous radiation from the amplifying medium during the FWM process. As a result of this process, two-mode squeezed light is generated in the sidebands within the range determined by the inverse relaxation time of the inversion of the medium $1/\tau_e$ ($\tau_e\sim 0.1-1$ ns). The optimal intensity of the strong wave is $I_{\text{sat}}\leq I \leq 2I_{\text{sat}}$. The degree of squeezing increases monotonously with increasing gain. Calculations predict about 10 dB of squeezing in the range $\leq 1$ GHz at optimal input intensity and unsaturated gain 15-20 dB. 

Thanks to their small size, low power consumption, and capability for direct integration onto photonic chips, semiconductor amplifiers are becoming reliable components for compact sources of squeezed light. 
The presented approach utilizing a semiconductor amplifier not only demonstrates ease of implementation but also meets the growing demands of modern scientific and industrial applications, thereby fostering the development of quantum technologies.

The four-wave mixing effect in the gain medium of semiconductor lasers suggests the possibility of generating light in a squeezed quantum state within the laser output.

\section{Acknowledgments}
We thank Nikolay A. Gippius, Dmitry A. Chermoshentsev and  Igor A. Bilenko for stimulating and useful discussions. 
The research was supported by the Russian Science Foundation, grant No. 26-72-10176.

\bibliography{sample}
\end{document}